\documentclass[showpacs,preprint]{revtex4b4}
\usepackage{epsfig}
\begin{document}

\title{
Variational approach to a class of nonlinear oscillators with several limit cycles}
\author{ M.\ C.\ Depassier}
\author{J. Mura} 
\affiliation{
 Facultad de F\'\i sica\\
	Pontificia Universidad Cat\'olica de Chile\\
	       Casilla 306, Santiago 22, Chile}

\date{\today}
%\maketitle
\begin{abstract}
We study limit cycles of nonlinear oscillators described by the equation  $\ddot x + \nu F(\dot x)  + x =0$ with $F$ an odd function. Depending on the nonlinearity, this equation may exhibit different number of limit cycles. We show that limit cycles correspond to relative extrema of a certain functional.  Analytical results in the limits $\nu \rightarrow 0$  and $\nu \rightarrow \infty$ 
are in agreement with previously known criteria.  For intermediate  $\nu$ numerical determination of the limit cycles can be obtained.
\end{abstract}

 
\maketitle

\section{Introduction}
The study of Lienard's differential equation 
\begin{equation}
\ddot x + \nu f(x) \dot x + x =0
\end{equation}
or the equivalent form, given earlier by Rayleigh,
\begin{equation}
\ddot z + \nu F(\dot z) + z = 0 \label{Rayleigh} \end{equation}
goes back  to the work of Rayleigh \cite{Rayleigh} motivated by  his work in acoustics and to  E. and H. Cartan \cite{Cartan}, van der Pol \cite{vdp} and Li\'enard \cite{Lienard} motivated by their work in electrical circuits. This type of equation arises directly in numerous applications and others can be reduced to them. The problem of studying the number and location of the periodic solutions for polynomial $F$  is a particular case of the general two dimensional problem $\dot x = P(x,y)$, $\dot y = Q(x,y)$, for polynomial $P$ and $Q$, which constitutes Hilbert's 16th problem. The problem we address here is the determination of the number and position of limit cycles of the equations above for even $f$, or equivalently, in Rayleigh's form for odd $F$. 
The two equations have the same number of limit cycles, Li\'enard's form (1) is obtained by taking the derivative of (2) and calling $\dot z = x$. These equations have a unique equilibrium point $z=0$ around which their periodic solutions will be nested. One of the most studied equations in this class is van der Pol's equation which has a unique limit cycle. Li\'enard \cite{Lienard} established conditions on $F$  which guarantee the existence of a single limit cycle. Many results on existence and uniqueness have been established, less is known about the number and location of limit cycles which do not satisfy Li\'enard's conditions 
\cite{Andronov,Qian,Perko}. Lins, de Melo and Pugh \cite{Lins} conjectured that if $F$ is a polynomial of degree $2 n + 1$ or $2 n + 2$ then there can be at most $n$ limit cycles.This conjecture was proved~\cite{Blows} for small $\nu$, that is, for small departures from the hamiltonian case. For $\nu \rightarrow 0$ the number and position of the limit cycles is given by the real roots of a polynomial obtained from Melnikov's function~\cite{Andronov,Perko}. In the present case  it reduces to finding the roots of 
\begin{equation}
\oint_{\Gamma_0} F[\dot z(t)] dt = 0 \label{Melnikov}
\end{equation}
where the integral is performed over a periodic solution $\Gamma_0$ of the hamiltonian system $\ddot z + z =0$, a circle in phase space. The conjecture has not been proved for arbitrary values of $\nu$ except for special cases. It holds true if $F$ is a polynomial of degree 3 or 4~\cite{Zeng} and if $F$ is odd of degree 5~\cite{Rychkov}. These results state the maximum number possible of limit cycles but not the exact number. Precise results on the exact number exist for particular cases, see for example~\cite{Figueiredo}.  In recent work the
case $\nu \rightarrow \infty$ has been studied, a criterion to determine the number of limit cycles in this asymptotic regime was given and tested in several examples~\cite{Lopez-Ruiz,Lopez}.
In a different approach, an algorithm to determine the number and position of the limit cycles for all values of $\nu$ and odd $F$ has been formulated. It is non perturbative in nature, it is based on finding the roots of a certain sequence of polynomials \cite{Giacomini1, Giacomini2,Llibre}. 
Fewer results~\cite{Cristopher} are known for the generalized equation $
\ddot x + \nu f(x) \dot x + g(x) =0.
$

In the present work we study limit cycles of (2), with $F(\dot z)$ and odd function. We show that limit cycles correspond to extrema of a certain functional. For small $\nu$ we recover known analytic results namely, (3); for $\nu \rightarrow \infty$ we recover  the results of~\cite{Lopez-Ruiz,Lopez}. For intermediate $\nu$ we must resort to numerical calculations.  The approach is a generalization of a method developed for other nonlinear problems \cite{BD1,BD2,BD3}. This first  approach  enabled us to show that in cases of cubic $F$ with a unique limit cycle, its location can be obtained from a minimum principle~\cite{BD4}.   Here we show that all limit cycles correspond to extrema of a certain functional which would allow us, at least numerically,  to count the number of limit cycles of a given equation. In Section 2 we derive the variational principle, in Section 3 we obtain analytically criteria for the small and large $\nu$ regime. In Section 4  we present some examples where we see that  approximate numerical determination of the limit cycles can be obtained  far from the asymptotic regimes. Concluding remarks are made in Section 5.

\section{Variational principle}

For odd $F$, due to the symmetry of  equation (2), the limit cycle extends between a minimum $z_{min} = -A$ and a maximum $z_{max} = A$. Moreover, in phase space,  if the point  $(\dot z, z)$ belongs to the limit cycle, then the point $(-\dot z,- z)$ also belongs to it. Therefore we may consider  the positive upper half $\dot z >0$ of the phase plane, where half a period will elapse when going from the points $(\dot z =0,z_{min})$ to   $(\dot z =0,z_{max})$.
Then the equation for the limit cycle in phase space $(\dot z(z),z)$ 
can be written as the nonlinear eigenvalue problem,
$
p  p_z + \nu F(p) + z = 0 $, 
with
$p(\pm A)  = 0$ and $p>0$.
Here we have called  $p(z) = \dot z (z)$ and the subscript denotes derivative. The eigenvalue is the amplitude $A$ which appears in the boundary conditions. It is convenient to define a new variable $u = z/A$ in terms of which the equation is 
\begin{equation}
{1\over S}\,p \frac{d p}{d u} +  F(p)+  R u = 0, \mbox{with}\quad
p(\pm 1) = 0, \quad p>0.
\label{eq:phase}
\end{equation}
Two parameters appear naturally, $R = A/\nu$ and $S = \nu A$. We may now construct the variational principle. Consider the extrema of the functional $R[p,\phi]$ of two arbitrary functions, $p(u)$ with $p(\pm 1)=0$, and $\phi(u)$, given by 
\begin{eqnarray}
R[p(u),\phi(u)] &=&
\frac
{\int_{-1}^1 \left( -{1\over S} p {d p\over d u}- F(p)\right)\phi(u)\,d u}
{\int_{-1}^1 u \phi(u) d u} \\
  &=&\frac
{\int_{-1}^1 \left[{1\over 2 S} p^2 {d\phi\over d u} - F(p) \phi(u)\right]d u }
{\int_{-1}^1 u \phi(u)\, d u}
 \label{integral} \end{eqnarray}
where the second expression is obtained after integration by parts.
Variation with respect to $p(u)$ at fixed $\phi(u)$ yields the equation
\begin{equation}
{1\over S} p {d\phi \over d u} - F_p \phi = 0.
\label{eq:cond1}
\end{equation}
Here $F_p$ denotes derivative of $F$ with respect to $p$. Variation with respect to $\phi$ at fixed $p$ yields
\begin{equation}
{1\over S} p {d p \over d u} + F(p) + R u  = 0,
\label{eq:cond2}
\end{equation}
that is, the equation for the limit cycles. Extrema of $R[p,\phi]$ satisfy both equations (\ref{eq:cond1}) and (\ref{eq:cond2}). Now  notice that equation (\ref{eq:cond1}) can be solved for $\phi$ in terms of $p$. Its solution is 
\begin{equation}
\phi(u) = \mbox{Exp}\left(  S \int_{-1}^u {F_p(p(t))\over p(t)} d t\right),
\label{eq:phi}
\end{equation}
Finally, replacing (\ref{eq:cond1}) in (\ref{integral}) we  obtain the main result, solutions of the equation for the limit cycles are extrema of
\begin{equation}
R[p] = \mbox{ext} \frac{\int_{-1}^1 \phi(u) \left({1\over2}p(u) F_p[p(u)]- F[p(u)]\right) d u}{\int_{-1}^1 \phi(u) u d u} \label{main}
\end{equation}
where the extremum is taken over all positive functions $p(u)$ that vanish at the end points and $\phi(u)$ is the function of $p(u)$ given by (\ref{eq:phi}). If we succeed in finding all the extrema, we have found all limit cycles.

\section{Small and Large $\nu$ Limits}

\subsection{Small $\nu$}

Here we shall see that for small $\nu$ we recover Melnikov's criterion.
From the definition of the parameters $R$ and $S$, we see that small or large $\nu$, for arbitrary $A$, corresponds to small or large $S$ respectively.  For small $\nu$ or equivalently for small $S$,
$$
\phi \approx 1 + S \int_{-1}^u {F_p[p(t)]\over p(t)}\, d t
$$
and
\begin{equation}
R \approx \mbox{ext} 
\frac
{\int_{-1}^1 ({1\over 2} p F_p - F)\, d u}
{S \int_{-1}^1 {1\over 2}  {F_p\over p} (1 - u^2)\, d u}.
\label{snu}
\end{equation}
Let us calculate the first variation $\delta R$ of R with respect to p.
We obtain 
\begin{equation}
S (R[p+ \delta p]-R[p])  = {1\over 2 D} \int_{-1}^1 \delta p \,(p F_{pp} - F_p)\left[ 1 - R S {(1-u^2)\over p}\right]
\label{variacion}
\end{equation}
where we called D the integral in the denominator of (\ref{snu}).  The term $p F_{pp} - F_p$ does not vanish. Then $\delta R =0$ for arbitrary $\delta p$ if 
$$
p(u) = R S (1-u^2).
$$
We know that this is the correct answer from direct integration of the equation.
In the small $\nu$ limit the cycle is approximately a solution of the equation ${1\over S} p {d p\over d u} + R u = 0$, with $p(\pm 1 ) =0$ whose solution is what we just obtained, 
the ellipse  $p(u) = R S  \sqrt{1-u^2}$.
This indicates that we should use a trial function of the form $p(u) = K \sqrt{1-u^2}$ and search for the value of $K$ for which $R$ has an extremum. We first notice that we will find some false extrema which we must discard. To see this, observe that with this trial function, $\delta p = \delta K \sqrt{1-u^2}$, and equation (\ref{variacion}) becomes
$$
S \delta R = {1\over 2 D} (1 - {R S\over K}) \int_{-1}^1 \delta K {p\over K} (p F_{pp} - F_p) d u,$$ and the integral will vanish for some $K$. This is not the desired solution, it is false and  obtained for not having swept over all possible trial functions.
With this in mind, we go back to (\ref{snu}). 
With $p(u) = A \sqrt{1-u^2}$ as the trial function, equation (\ref{snu}) is of the form
$$
R \approx \mbox{ext} {A^2\over S} {h(A)\over g(A)}
$$
where we called
$$ h(A) = \int_{-1}^1 ({1\over 2} p F_p - F) d u, $$
$$ g(A) = \int_{-1}^1 {1\over 2} p F_p  d u.$$
It is easy to verify that $A h' - A g' = - 2 g$, and we obtain
$$
{d R \over d A} = \frac{2 g h + A (h' g - h g')}{S g^2} = \frac{A h'(A) (h - g)}{S g^2}.
$$
Extrema of $R$ occur when 
$A h'(A)  =0$   and when $h(A) = g(A)$. The first condition is either $A=0$, the trivial solution which is always present or when $h'(A) =0$. This solution is precisely the false solution which we discard.  We retain then the solution $h(A) =g(A)$, which is simply
$$  \int_{-1}^1  F[p(u)]) d u = 0,  \qquad \mbox{with}\qquad p(u) = A \sqrt{1-u^2}.
$$
This is exactly condition (\ref{Melnikov}) since we have considered only odd $F$. 

\subsection{Large $\nu$}

In recent work L\'opez et al.\cite{Lopez} study limit cycles of Li\'enard's differential equation (1) in the strongly nonlinear regime. Their approach is based in constructing approximate solutions to the differential equation. Here we give a heuristic derivation of their result. Since $S$ appears only in the exponential in the form Exp(S z(u)), we know, from Watson's lemma \cite{Bender}, that when $S \rightarrow \infty$  the leading contribution to the integral comes from the points where the term in the exponential has an extremum. Here the term in the exponential is 
$$
z(u) = \int_{-1}^u {F_p(p(t)\over p(t)} d t.
$$
Extrema occur where $z'(u)$ = 0, that is where $F_p = 0$. For large $S$ then, $R$ will be given 
by
$$
R = \mbox{ext} \left( {-F(\hat p)\over \hat u}\right),$$
where $\hat p$ is the solution of $F_p (\hat p) = 0$ and $\hat u$ is the value of $u$ for which $p = \hat p$ on the orbit.  The extremum is now taken over $\hat u$. Since $R$ is positive,
possible extrema  of $R$ will occur at $\hat u = 1$ if $F(\hat p) < 0$ or at $\hat u = -1$ if $F(\hat p) > 0$. The value of $R$ at the extremum points, if any,  for large $S$, are given then by
$$
R = |F(\hat p_1)|, \quad |F(\hat p_2)|, \, \mbox{etc}.
$$
In order to see which of these correspond to true extrema, hence to limit cycles, we return to the differential equation. For large $S$, limit cycles are approximately given by, 
\begin{equation}
F(p) + R u = 0
\label{asint}
\end{equation}
the solution of which has to be matched to a thin boundary layer.
Now that we know the possible values of $R$, we can read directly which are the limit cycles. This is best seen in a plot in phase space of (\ref{asint}). Suppose there are several points $\hat p$  at which $F_p =0$. Since $F$ is odd, take the positive values and label them in growing order, $ p_1 <  p_2 <  p_3 ...$. Correspondingly we have different values of $R$. Since limit cycles must be nested, and, for the systems considered, the derivative $d p/d u$ can vanish only at one point, the only limit cycles will be those for which $|F( p_j)| > |F(p_k)|$ for all $k < j$.  This is best seen in Fig. 1 where we show a case with two allowed limit cycles. In the asymptotic regime when $u\rightarrow \pm 1$ the horizontal coordinate tends to $R = F(\hat p)$. The trajectory of each limit cycle is indicated by the arrows. From the definition of $R$ we know then that in the limit $\nu \rightarrow \infty$, for each allowed cycle, the amplitude grows as $A_i = |F( p_i)| \nu$. Moreover we can read the maximum value of $p$ in each cycle. The maximum $p$ are solutions of  $F(p) + |F(\hat p)| = 0$. The values of $p$ determined in this way corresponds to the amplitude of the associated Li\'enard equation (1).  Thus we  have recovered the solution of \cite{Lopez,Lopez-Ruiz}, which gives support to  the conjecture of Lins, de Melo and Pugh.

\section{Examples}

Having seen that we recover the small $\nu$ and large $\nu$ limits,  we now give numerical results for  arbitrary values of $\nu$ in simple examples. 
Here too, false extrema may appear due to the impossibility of sweeping over all trial functions. This becomes evident, as in the $\nu \rightarrow 0$ limit, by considering the first variation of $R$ with respect to $p$, we find that for arbitrary $S$, 
$
R(p +\delta p) - R(p)$
vanishes when
$$
\int_{-1}^1 du \,\delta p \left[{F_{pp}\over p} -{F_p\over p^2}\right]
\left\{
{1\over2}p(u)^2 \phi(u) + S \int_{-1}^u d t \, \phi(t) 
\left[ R t - \left( {1\over 2} p(t) F_p(p(t))  - F(p(t))\right)\right]
\right\} = 0.
$$
As in the small $\nu$ limit, the term in square brackets does not vanish, so 
for arbitrary $\delta p$ the first variation vanishes when the term in curly brackets vanishes.
Taking the derivative of this term one can see that it corresponds to the equation for the limit cycles. In practice though,  when sweeping over a restricted variety of trial functions the integral may vanish at other points, which calls for some care.
To illustrate the use of the variational principle we shall use for all $\nu$ a simple  trial function with only one parameter, the  ellipse $p(u) = K \sqrt{1-u^2}$. For each value of $S$ we insert this trial function in (\ref{main}) and sweep in $K$ to find all the extrema. We begin with small $S$ where we identify the true minima and follow their evolution as $S$ is increased. We obtain a table of extrema $(R_1, R_2, R_3, ...)$ for each $S$, from which we compute the values  
$\nu_i = \sqrt{S/R_i}$ and the corresponding amplitude $A_i =\sqrt{S R_i}$. As we show below with this very simple trial function, appropriate for $\nu \rightarrow 0$, we obtain the approximate location of the limit cycles for $\nu$ well beyond the region of validity of this trial function.

As a first example we take 
$$
F(p) = {4\over 5}p - {4\over 3} p^3 + {8\over 25} p^5.$$
For $\nu \rightarrow 0$, (\ref{Melnikov}) predicts the existence of two limit cycles, of amplitudes $A=1$ and $A=2$. As $\nu$ increases the amplitudes change. In Fig. 2 we show the amplitudes as functions of $\nu$. The continuous line is the amplitude obtained from the direct integration of the differential equation (\ref{Rayleigh}). Unstable limit cycles are obtained by integrating the equation with $\nu < 0$ since this only changes the stability properties of the cycles.  The origin is a stable fixed point, the inner limit cycle unstable and the outer stable. The dots indicate the values of the amplitude obtained following the extrema of $R[p]$. The agreement is close, in spite of having used a one parameter trial function.

A more interesting example is provided by
$$
F(p) = p -\sqrt{{41\over 9}} p^3 + p^5
$$
Here (3) again predicts the existence of two limit cycles for $\nu \rightarrow 0$ of amplitudes 
$A = \sqrt{(\sqrt{41}-1)/5} \approx  1.039$ and $A= \sqrt{(\sqrt{41}+1)/5}\approx 1.217$. For this equation it is known that for large $\nu$ there is no limit cycle. A  bound on the value of $\nu$ for which no limit cycles exist is known \cite{Alsholm}. The criterion for $\nu \rightarrow \infty$ also indicates that no limit cycles exist in this regime.  Numerical integration of the differential equation shows that, as  $\nu$ increases these two limit cycles merge and disappear. In Fig. 3 we show the amplitude as a function of $\nu$. The continuous line is obtained by direct numerical integration and the dots were obtained variationally. There is qualitative agreement, the variational result predicts the merging of the limit cycles. The exact cycles are very different from the circles we use as trial function so this trial function is not adequate at larger $\nu$. 

\section{Summary}

We have shown that all limit cycles of (\ref{Rayleigh}) correspond to extrema of a certain functional. The exact position of each limit cycle is obtained when the trial function coincides with the solution, otherwise an approximate estimation can be made. 
From the variational expression analytical results can be obtained in the two asymptotic limits, $\nu \rightarrow 0$ and $\nu \rightarrow \infty$. In these asymptotic regimes we obtain both the amplitudes of the Rayleigh and Lienard form of the equations.  In the small $\nu$ limit we reobtain the known criterion, namely Melnikov's integral. In the large $\nu$ regime our results coincide with that obtained in recent work. In the intermediate regime  the number and position 
of limit cycles can be obtained numerically. Even  with  simple trial functions  relatively close estimates are obtained. Whether it is possible to count analytically the number of extrema, hence of limit cycles, is a question that remains to be answered.

\begin{acknowledgments}
 We thank R. Benguria for helpful discussions. 
This work was supported by Fondecyt project 1990429 (Chile).
 \end{acknowledgments}

\begin{figure}[p]
\centering\epsfig{figure=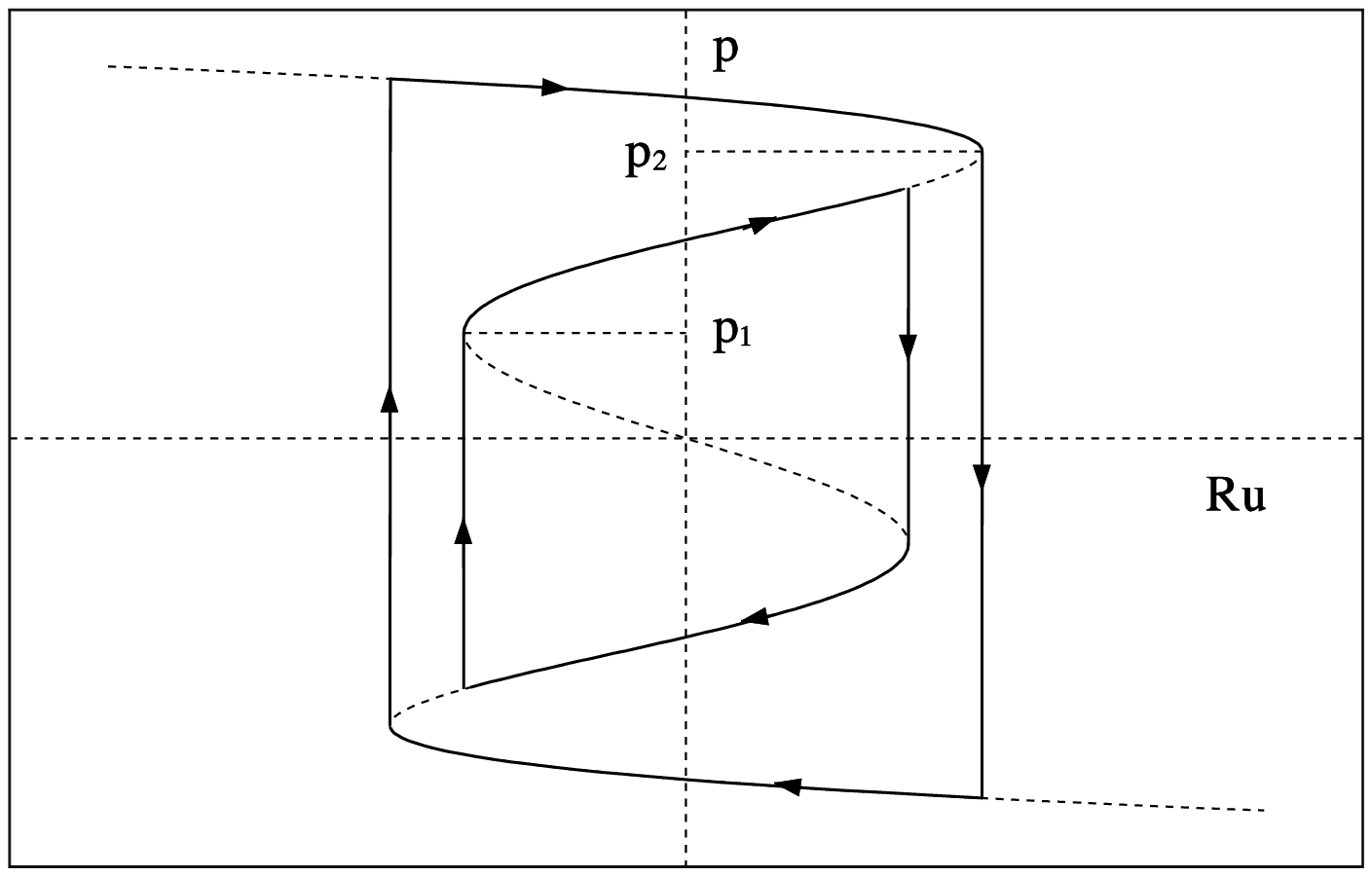,width=12.0truecm}
\caption{Graph of allowed limit cycles in the $\nu \rightarrow \infty$ regime.}
\end{figure}

\begin{figure}[p]
\centering\epsfig{figure=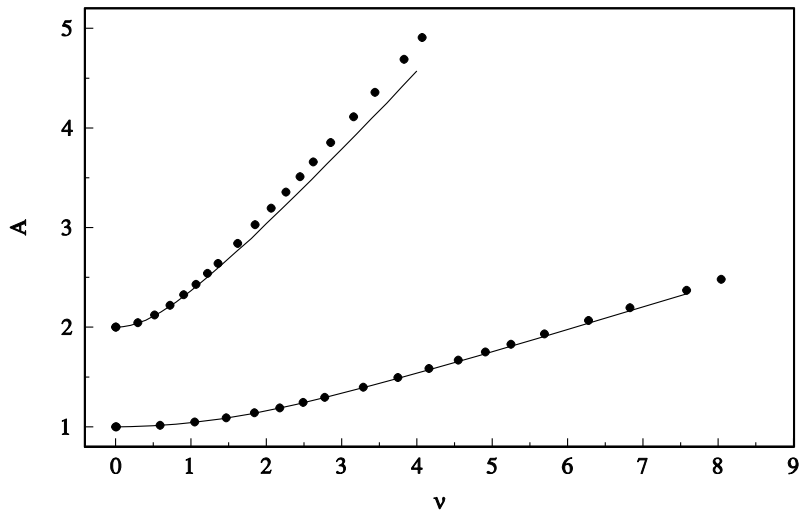,width=12.0truecm}
\caption{Amplitude of the limit cycles for $F(p) = (4/5) p - (4/ 3) p^3 + (8/ 25) p^5$. The continuous line is the result of numerical integration of the differential equation. The dots were obtained variationally with a simple trial function.}
\end{figure}

\begin{figure}[p]
\centering\epsfig{figure=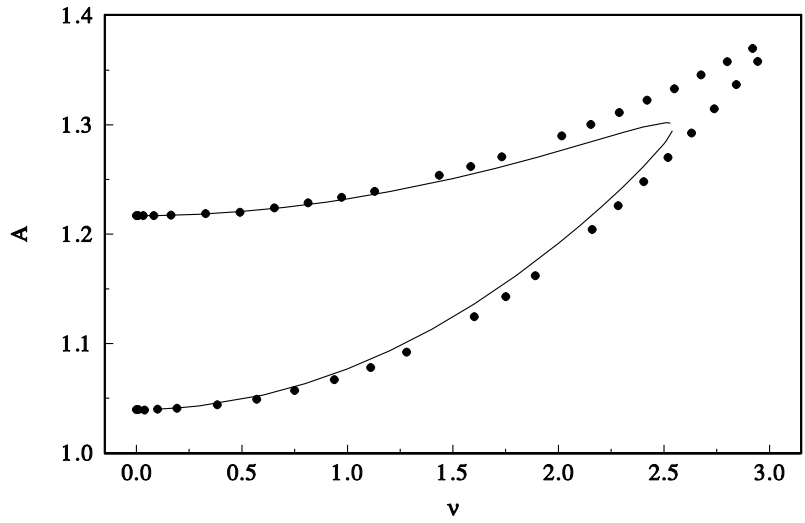,width=12.0truecm}
\caption{As in Fig.2, for $F(p) = p -\sqrt{41/ 9} p^3 + p^5$.}
\end{figure}

\end{document}